# Dual-band polarized upconversion photoluminescence enhanced by resonant dielectric metasurfaces

Ziwei Feng[1,2†], Tan Shi[1†], Guangzhou Geng[3†], Junjie Li[3*], Zi-Lan Deng[1*], Yuri Kivshar[4,5*], and Xiangping Li[1*]

[1]*Guangdong Provincial Key Laboratory of Optical Fiber Sensing and Communications, Institute of Photonics Technology, Jinan University, Guangzhou 511443, China.*
[2]*Wuhan National Laboratory for Optoelectronics, Huazhong University of Science and Technology, Wuhan 430074, China.*
[3]*Beijing National Laboratory for Condensed Matter Physics, Institute of Physics, Chinese Academy of Sciences, Beijing 100190, China*
[4]*Nonlinear Physics Centre, Research School of Physics, Australian National University, Canberra ACT 2601, Australia*
[5]*Qingdao Innovation and Development Base of Harbin Engineering University, Qingdao 266000, China*

*Corresponding authors: yuri.kivshar@anu.edu.au; xiangpingli@jnu.edu.cn; zilandeng@jnu.edu.cn; jjli@iphy.ac.cn
†These authors contributed equally to this work.

**Abstract**

Lanthanide-doped upconversion nanoparticles emerged recently as an attractive material platform underpinning a broad range of innovative applications such as optical cryptography, luminescent probes, and lasing. However, the intricate 4f-associated electronic transition in upconversion nanoparticles leads only to a weak photoluminescence intensity and unpolarized emission, hindering many applications that demand ultrabright and polarized light sources. Here, we uncover a new strategy for achieving ultrabright and dual-band polarized upconversion photoluminescence. We employ resonant dielectric metasurfaces supporting high-quality resonant modes at dual upconversion bands enabling two-order-of-magnitude amplification of upconversion emissions. We demonstrate that dual-band resonances can be selectively switched on polarization, endowing cross-polarization controlled upconversion luminescence with ultra-high degrees of polarization, reaching approximately 0.86 and 0.91 at dual emission wavelengths of 540 nm and 660 nm, respectively. Our strategy offers an effective approach for enhancing photon upconversion processes paving the way towards efficient low-threshold polarization upconversion lasers.

**Introduction**

Lanthanide-doped upconversion nanoparticles (UCNPs) demonstrate many superior properties of broadly-tunable multicolor emission and long emission lifetimes making them well-suited for many applications including super-resolution imaging[1-5], optical multiplexing[6, 7], three-dimensional displays[8] and low-threshold lasing[9, 10]. Despite such advantages, the further progress and practical applications of UCNPs face significant challenges due to low upconversion efficiencies and a lack of polarization emission control. Polarization, as another important characteristic of fluorescence, provides orientation and structural information in an additional dimension, and it has been extensively employed in the fluorescence polarization imaging technique[11, 12]. Unfortunately, the inherent upconversion processes of UCNPs, which are associated with the 4f electronic transition, are typically weak and lacks polarization properties[13].

Previous studies have focused primarily on enhancing the upconversion luminescence by coupling UCNPs with plasmonic nanostructures[14-21]. The concentrated electromagnetic field in the vicinity of these nanostructures can significantly boost both the absorption cross-sections and the upconversion radiation rates. In the meantime, it has been revealed that the luminescence's intensity and polarization state can be remarkably tailored by the localized plasmonic resonances[17, 19, 22]. However, these studies have concentrated on the polarization properties of the overall luminescence, which cannot achieve selective modulation of polarization characteristics for multi-color upconversion bands. Moreover, the substantial intrinsic Ohmic loss in metallic nanostructures leads to plasmonic resonances with generally low values of the quality factors (Q factors) and limited enhancements. In addition, a contact between metal and upconversion nanoparticles may introduce undesirable quenching effects and even compromise their performance.

In contrast to plasmonic nanostructures, all-dielectric resonant metasurfaces composed of high-index dielectric nanostructures with low losses at visible frequencies support both electrical and magnetic multipolar Mie-resonant responses[23-27]. Furthermore, collective high-Q resonances such as lattice resonances and bound states in the continuum (BICs) emerge due to a coupling of these Mie multipoles in periodic arrays[28-31]. Among them, BICs, which are completely decoupled from radiative modes, have recently attracted considerable attention due to their infinitely-high Q factors[30, 32]. Breaking symmetry of meta-atoms can transform a BIC into a quasi-BIC with finite but extraordinarily high Q factors[33-36]. These emerging concepts have recently been well-exploited and demonstrated for fluorescence enhancement and low-threshold lasing applications[37-39].

Here, we utilize the emerging concept of high-Q collective modes of resonant metasurfaces for polarization-controlled dual-band upconversion burst as illustrated in Figure 1. We design and fabricate high-Q resonant metasurfaces composed of Titanium dioxide ($TiO_2$) diatomic nanobricks (Figs. 1a and 1b), which support a quasi-BIC mode (with Q~1031) at the wavelength of 660 nm for the $y$-polarized incidence, and another high-Q Mie resonance mode (with Q~1044) at the wavelength of 540 nm for the $x$-polarized incidence, respectively. Consequently, $NaYF_4$:Yb/Er UCNPs deposited on the metasurface exhibit ultrabright upconversion luminescence at dual bands of 540 nm and 660 nm, with enhancement factors up to approximately 51-fold and 43-fold. By controlling the rotation of the polarization analyzer, we demonstrate linearly- and cross-polarized emission at dual bands with ultra-high degrees of polarization (DoPs). Our demonstration paves the way for efficient enhancements and polarization control of the UCNP emission with potential applications in low-threshold polarization upconversion lasers and hyperspectral imaging/sensing.

**Results and discussion**

The unit-cell of the proposed dielectric resonant metasurface is shown in Figure 1b. It is constructed by a pair of $TiO_2$ (refractive indexes refer to Supplementary Fig. S1) nanobricks placed on a $SiO_2$ substrate. The nanobricks are of height (h) 450 nm, length ($w_y$) 240 nm and width ($w_x$) 100 nm. The periods along the $x$- and $y$-directions are $p_x$ = 430 nm, and $p_y$ = 345 nm, respectively. A BIC can be formed in such metasurface without any radiation due to symmetry protection or destructive interference between different radiation channels. In our design, the introduction of the asymmetric parameter $\delta$ (the orientation angle $\delta$ of the nanobrick with respect to the $y$ axis) breaks the rotational symmetry protection of BICs, transferring them into high Q-factor quasi-BICs that are accessible with free-space excitations at normal incidence. By varying the orientation angle between the two nanobricks and the background refractive index, we can modulate resonant wavelength to match the emission wavelength of the UCNPs. As depicted in Figs. 1c–e, $NaYF_4$:Yb/Er (20/2 mol %) nanoparticles with size of around 15 nm (Supplementary Fig. S2) were selected to be spin-coated onto the metasurface to adjust the background refractive index. The designed metasurface exhibits a quasi-BIC mode at λ ≈ 660 nm for $y$-polarization, and another Mie resonance mode at λ ≈ 540 nm for *x*-polarization light. Those two high-Q resonance modes overlap with the emission band of the $Er^{3+}$ upconversion transition ($^4H_{11/2}$ to $^4I_{15/2}$ and $^4F_{9/2}$ to $^4I_{15/2}$).

To further get insights into the resonance behavior, we calculate the reflectance spectra of the metasurface with respect to asymmetry parameter *δ* under *x*- and *y*- polarized normal incidence (Figs. 2a-b). Under *y*-polarized normal incidence, the metasurface supports a symmetry-protected BIC at *δ*=0º, which has an infinite Q-factor indicated by the red circle. A slight symmetry breaking (*δ*≠0 º) can transfer the symmetry-protected BIC into a quasi-BIC. Supplementary Fig. S3 shows the Q-factor and the resonance wavelength of the quasi-BIC mode on different asymmetry parameter *δ*. As the *δ* increases, the Q factor remains high and the resonance wavelength red-shifts. A spectrally narrow resonance at ∼660 nm can be observed for the case of *δ* = 10°, in which case the Q factor is calculated to be 1031.

For *x*-polarized normal incidence, multiple linewidth-vanishing BIC modes are identified in Figure 2b, with different resonant wavelengths. Among them, a Mie resonance at 540 nm sustains as indicated by the green circle, regardless of variations in the asymmetry parameter δ. Moreover, the linewidth and resonance wavelength keep almost intact during the variation, as shown in Figure 2b and Supplementary Fig. 3. We calculate the electromagnetic near field distributions and the multipole mode expansion spectra of the Mie resonance. It reveals that the Mie resonance is dominated by the toroidal (TD) mode, with smaller contributions from other dipole modes, considering the lattice interaction and the influence of periodicity (Supplementary Fig. S4). Supplementary Fig. S5 illustrates that variations in other structural parameters ($p_x$, $p_y$, $w_x$, $w_y$, $h$ and *n*) only affect the peak position of the Mie resonance. To match the emission wavelengths of the UCNPs, we judiciously adjust the asymmetric parameters to introduce high-Q resonances at 540 nm and 660 nm, respectively, responding to *x*- and *y*-polarizations. Along with resonance wavelength and polarization tunability, both the high-Q quasi-BIC and the Mie resonance mode strongly amplify the electric near fields around the resonators with intensity enhancement factors $I/I_0$ up to 1600 (Figures 2c-d). This enables a strong confinement of the radiative emission field inside the metasurface, which exhibits a high field intensity for efficient upconversion luminescence enhancement.

The metasurface was fabricated using electron beam lithography (EBL) technology. Figure 2e shows slant-view scanning electron microscopy (SEM) images of the fabricated sample.

NaYF$_4$:Yb/Er (20/2 mol %) nanoparticles were spin-coated onto the metasurface, as shown in the top-view SEM image (Figure 2f). The UCNPs are uniformly filled in TiO$_2$ nanobricks, which changes the background refractive index of the metasurface. The spectral measurement reveals that UCNPs/metasurface sample exhibit resonance peaks at ~540 nm and 660 nm under different polarizations, which perfectly overlap with the luminescence peaks of UCNPs (Figure 2g). The experimentally achieved Q-factor of the Mie resonance mode at 540 nm is 151, and that of the quasi-BIC mode is around 268. The smaller values of Q-factors compared with the theoretical ones are due to the finite array size and fabrication imperfections.

We further experimentally validated the feasibility of our design to achieve polarization-controlled upconversion emission enhancement. Figures 3a and 3b show the optical band structure of the resonant metasurface when the incident light is along the *x*- and *y*-direction, respectively. We note that *x*- and *y*-polarization incidences show a clear response difference at the Γ point. The *x*-polarization incidences show strong reflection while *y*-polarization excitation shows nearly zero reflection at 540 nm, as indicated by the white circle. In contrast, the strong reflection in *y*-polarization excitation happens at 660 nm. The polarized upconversion luminescence from the UCNPs/metasurface sample was then collected by using home-made confocal system with discrimination capability (Supplementary Fig.6). Under excitation at 980 nm, the luminescence spectra of UCNPs on metasurface and glass are shown in Figures 3c-d. Er$^{3+}$-activated UCNPs showed two characteristic bands centered around 540 and 660 nm. Compared to the luminescence intensity on the glass substrate, the UCNPs/metasurface sample exhibit sharp green emission in *x*-polarization detection. On the contrary, the luminescence intensity at 660 nm is substantially enhanced in *y*-polarization detection than that on glass. In the meantime, the Fourier-plane upconversion luminescence images at 540 nm and 660 nm were measured using a bandpass filter are shown in insets of Figures 3c-d. The upconversion luminescence images display characteristic angle dispersion at dual bands, which implies that emissions of UCNPs are polarized and coupled to Mie resonance mode and the quasi-BIC mode of the resonant metasurface, respectively.

We found that the linewidth of fluorescence luminescence was significantly narrower compared to that on glass substrate. Figure 3e displays the statistical chart of the linewidths of the upconversion emission for UCNPs on the metasurface and on the glass substrate. The luminescence bandwidths at emission wavelengths 540 nm and 660 nm on the metasurface are 3.18 nm and 3.67 nm, respectively, which are close to the reflection spectrum bandwidth of the metasurface with UCNPs coating. In comparison to the upconversion fluorescence bandwidth on the glass substrate, the bandwidth on the metasurface is reduced by approximately four times. This finding offers new insights into narrowing linewidth of fluorescence emission for increasing fluorescence detection sensitivity and reducing background noise. Figure 3f illustrates the excitation power dependence of the UCNPs coupled metasurface and the reference samples. At low power excitation, both green and red emissions are present with slopes of 2.2 and 2.3, respectively, indicating that the upconversion emission is basically a two-photon process. It was observed that luminescence intensity increases linearly with incident power under strong excitation. However, the UCNPs coupled with the resonant metasurface exhibit a smaller slope and significant fluorescence enhancement due to the local field enhancement of the quasi-BIC mode and the Mie resonant mode. The upconversion emission rapidly reaches a saturation state, resulting in a reduced slope. Moreover, we quantified the enhancement factor for the upconversion luminescence intensity of each emission band under different excitation intensities (Figure 3g). In high-power excitation, the enhancement factor is 7-fold for 540 nm and 5-fold for 660 nm, while it increases to 51-fold and 43-fold as the incident power is reduced to 92 $Wcm^{-2}$.

The physical mechanism behind the enhancement of upconversion fluorescence stems from the near-field enhancement induced by the quasi-BIC and Mie-resonant modes. The alterations in the electromagnetic field near the UCNPs may result in an excited flux, while the increased local density of optical states (LDOS) offers additional decay channels for the UCNPs to radiate via the quasi-BIC and Mie resonant modes into the far-field. To characterize the enhanced spontaneous-emission processes, we measured the radiative lifetime for the UCNPs-BIC sample and UCNPs on the reference glass (Figures 3h-i). UCNPs on the resonant metasurface exhibit a decay rate enhancement of 3.2 at 540 nm. Simultaneously, the enhancement of the decay rate is 2.7 times greater than the glass sample at 660 nm.

The enhancement of the radiative decay rate of the resonant metasurface can also be explained through Purcell factor analysis, which was investigated by simulating a finite resonant metasurface with 30 × 30 unit-cells, consistent with the illumination area in the experiment, where an objective with a numerical aperture (NA) of 0.4 was used for both excitation and light collection. The corresponding Purcell factor of different locations along the *x*-axis at 540 nm in *x*-polarization and 660 nm in *y*-polarization revealed the decay rate enhancement for red and green emissions, respectively (Supplementary Fig.7). It can be observed that the Purcell factor is very close to the changes in the emission decay rate. Furthermore, the Purcell factors at the electric field maxima at different positions are similar, suggesting that the quasi-BIC and Mie resonance with a large LDOS provide additional decay channels to enhance the decay rate and improve upconversion emission.

The inherent polarization-controlled high Q resonant modes in such resonant metasurface enables the selective manipulation of the dual-band emission polarizations. Figure 4a shows the polarized upconversion emission spectra of UCNPs on the metasurface, recorded by rotating an analyzer before the spectrometer. As the analyzer varies from 0° to 180°, the upconversion luminescence intensity of 540 nm and 660 nm on metasurface cross-change sinusoidally. The polar plot of the up-conversion luminescence intensity as a function of the polarization angle shows evident "8"-shaped convergence (Figure 4c). The luminescence intensity at 540 nm shows a distinct *x*-polarized emission with the weakest and strongest emission at 0 and 90°, respectively. Conversely, the red emission at 660 nm shows the opposite case, indicating their countercyclical polarization anisotropy. On the glass substrate, the intensity of UC luminescence remains almost unchanged, and an approximate circle is drawn in polar plates, indicating no polarization response. The degree of polarization (DoP) of the upconversion luminescence can be evaluated as DoP = $(I_{max} - I_{min})/(I_{max} + I_{min})$, where $I_{max}$ and $I_{min}$ represent the maximum and minimum values of upconversion luminescence intensity. The luminescence emission of UCNPs on the reference glass shows no sensitivity against to the polarization. In stark contrast, the luminescence emission DoP of UCNPs coupled with the resonant metasurface achieves ~0.86 and ~0.91 for emissions at 540 and 660 nm, respectively. The luminescence intensities of green and red are selectively enhanced for UCNPs on the metasurface, providing a unique approach to achieve luminescence polarization control in nanophotonic devices.

In summary, we have demonstrated the substantial augmentation and high polarization of upconversion luminescence from upconversion nanoparticles using all-dielectric resonant metasurfaces. Supporting polarization-controlled quasi-BIC and Mie modes with ultra-high Q factors at dual upconversion bands, such resonant metasurfaces provides an enabling platform for significantly amplified upconversion emissions, achieving 51- and 43-fold enhancements, as well as simultaneous cross-polarized luminescence with DoP reaching 0.86 and 0.91, for green and red emissions, respectively. We believe our results provide a new approach for a design of ultrabright and polarized nanoscale emission sources holding great potential for the further advances in upconversion polarization lasing, biosensing, and fluorescence imaging.

**Method**

**Fabrication of samples**. The metasurfaces were fabricated on a fused quartz substrate using a combination of processes, including electron beam lithography (EBL), atomic layer deposition (ALD), patterning, lift-off, and etching. First, the $TiO_2$ film was deposited by the atomic layer deposition method, during which the continuous 20 sccm flow of $N_2$ carrier gas and maintained at 105 °C. Then the polymethyl methacrylate (PMMA) resist 950-A7 was spun on fused quartz substrate and baked at 180 °C for 2 mins. The designed structures were patterned by using electron beam lithography system (6300FS, JEOL) and then developed in a mixture of MIBK and IPA with a ratio of 1:3. Finally, a dry etching process was performed in the ICP-RIE system (Plasmalab System 100 ICP180, Oxford) with a mixed reactive gas of $BCl_3$ and $Cl_2$ to remove the $TiO_2$ film on the top of the resist. And dry etching process with Oxygen also was used to remove the residual resist.

**Characterization of metasurface.** The broadband source (Fianium-WL-SC480) through a polarizer and half-wave plate generates different linear polarizations. The excitation beam is transmitted by beam splitters, then focused on the samples through an objective lens (NA=0.4). The reflection spectra of the resonant metasurface were measured using an Ocean spectrometer (USB4000).

**Characterization of luminescence spectra and lifetime**. Luminescence spectra were measured using a CW fiber-coupled laser diode ($\lambda_{ex}$ = 980 nm) as the excitation source. The objective lens with a numerical aperture (NA) of 0.4 was used for both excitation and emission collection. Then, the incident light was focused on the sample and the emission through polarizer and filter was captured by a spectrometer (Shamrock 303i, Andor) equipped with an electron-multiplying charge-coupled device camera (EMCCD, Newtown, Andor). And time-resolved photoluminescence measurements of UCNPs were performed by modulating the 980 nm laser with a signal generator.

The luminescence is collected by a single-photon avalanche diode (SPAD; SPCM-AQRH-14-FC) and synchronized with a modulated excitation signal by a data acquisition (DAQ).

**Contributions**

T.S. and Z.F. conceived the idea. Z.F, T.S., Z.-L. D., and X.L designed the experiments. T.S., and Z.F. carried out the design and simulation of the metasurfaces. Z.F., T.S., Z.-L. D., Y.K., and X.L. conducted the theoretical analysis of the results. G.G. and J.L. fabricated the samples. Z. F and T.S. performed the measurements. Z.-L. D., Y.K., and X.L. supervised the project. Z. F., T.S., Z.-L. D. and X.L. analyzed the data and wrote the paper. All authors participated in the discussion and analysis of the manuscript.

**Acknowledgement**

This work was supporting by national Key R&D Program of China (2021YFB2802003, 2022YFB3607300), the China Postdoctoral Science Foundation funded project (No.2022M711241), National Natural Science Foundation of China (NSFC) (62075084), and the Guangdong Basic and Applied Basic Research Foundation (2022B1515020004).

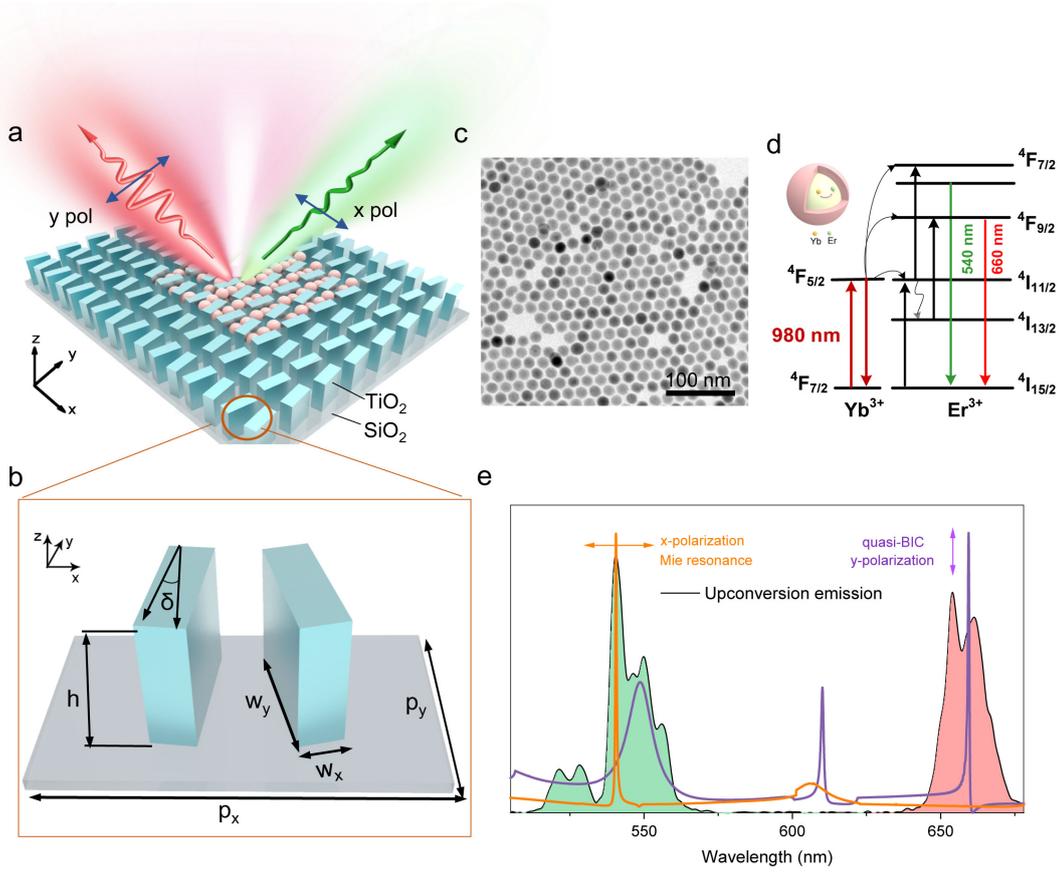

**Figure 1.** Polarization-controlled upconversion emission enhancement. (a) Schematic presentation of a resonant metasurface with high-Q quasi-BIC and Mie resonant modes. The metasurface is composed of an array of TiO$_2$ nanobricks loaded with UCNPs. (b) Diagram of the designed unit cell depicting dual TiO$_2$ nanobricks with a tilt angle of $\delta = \pm 10°$ relative to the $y$-axis, placed on a SiO$_2$ substrate. Geometrical parameters: $w_x$=100 nm, $w_y$ = 240 nm, $h$ = 450 nm. The rectangular lattices periods are $p_x$ = 430 nm and $p_y$ = 345 nm. (c) TEM image of NaYF$_4$:Yb/Er upconversion nanoparticles. (d) Energy level diagram illustrating upconversion processes of Yb$^{3+}$/, Er$^{3+}$. (d) Energy level diagram illustrating the upconversion transition of Yb$^{3+}$/Er$^{3+}$. (e) $x$- and $y$-polarization controlled reflection spectra of the resonant metasurfaces overlapping with the emission spectra of the NaYF$_4$:Yb/Er upconversion nanoparticles. Yellow curves denote Mie resonance under $x$-polarization and magenta curves denote quasi-BIC resonance under $y$-polarization incidence.

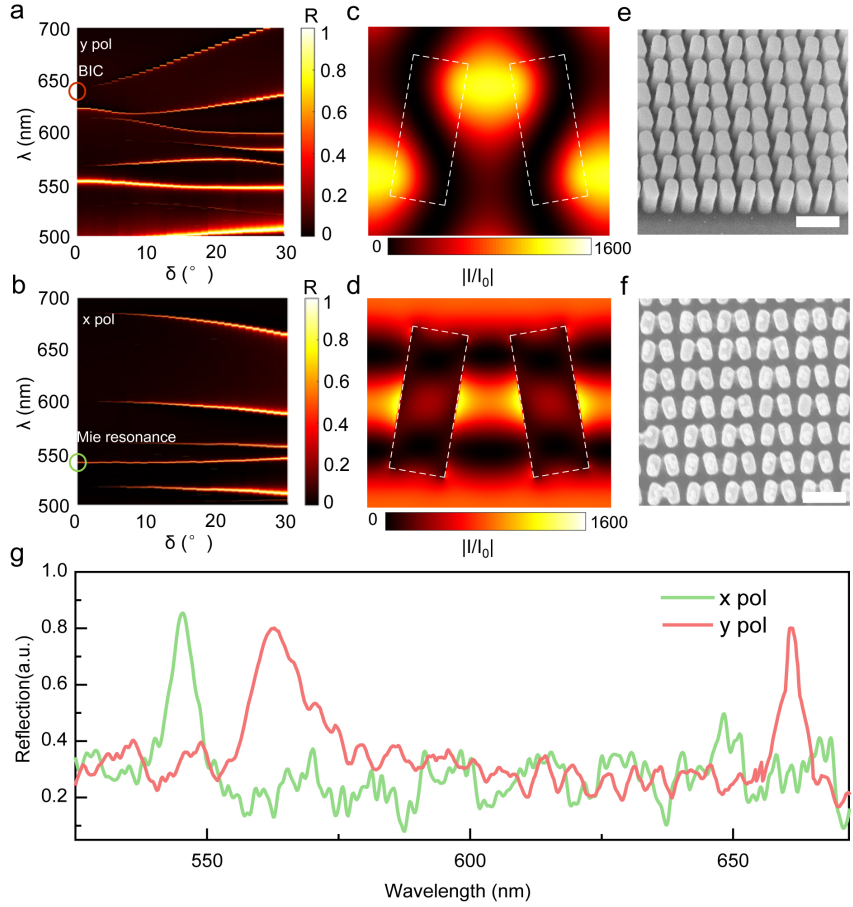

**Figure 2**. (a-b) The evolution of reflection spectra by continuously varying the geometry asymmetry parameter *δ* under *y*-polarization (a) and *x*-polarization (b). The vanishing point indicated by the red circle denotes the BIC mode. The green circle denotes the Mie resonance. (c-d) Electric field distributions at resonant wavelength 660 nm for *y* polarization, and 540 nm for *x* polarization. (e) Slant view SEM image of the resonant metasurface with background refractive index of 1. (f) Top view SEM image of the resonant metasurface loaded with UCNPs. Scale bar: 500 nm. (g) The measured reflection spectra of the unloaded metasurface under *x*- (green) and *y*- (red) polarizations, respectively.

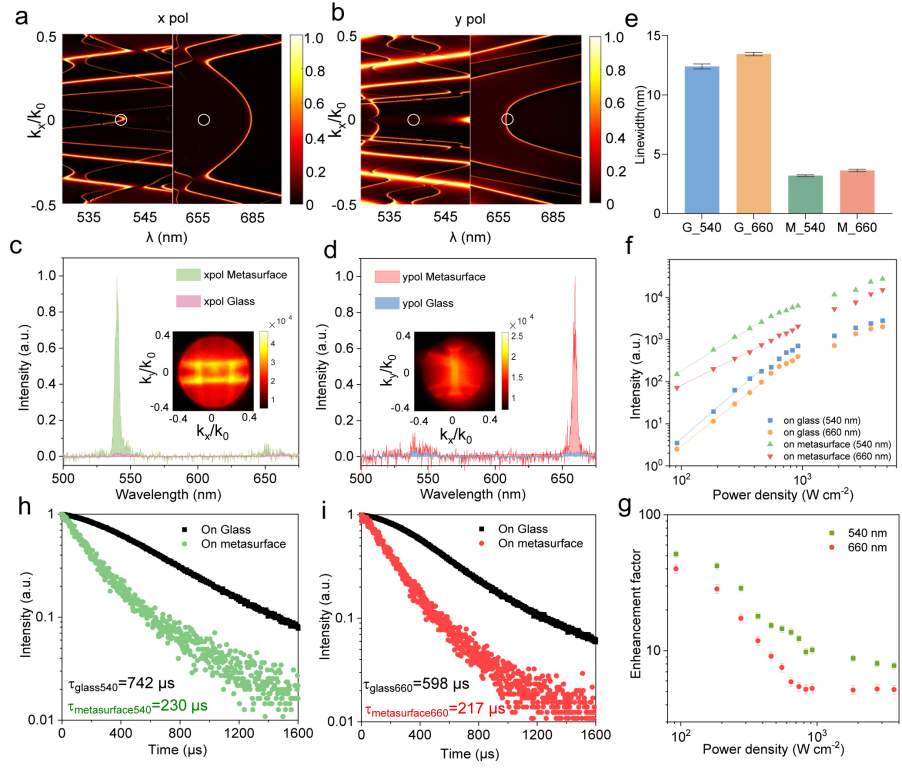

**Figure 3**. (a-b) Simulated angle-resolved reflectance spectra under (a) *x*-polarized beam and (b) *y*-polarized beam for the wavelength range 530-700 nm. The ratio between the projected in-plane momentum $k_x$ and free-space momentum $k_0$ indicates the reflection angle $\sin\theta = k_x/k_0$. (c-d) Upconversion luminescence spectra from UCNPs deposited on the metasurface and on glass substrate under *x*-polarization detection (c) and *y*-polarization detection (d). Inset: Fourier-plane upconversion photoluminescence images. (e)The linewidth of luminescence spectra of UCNPs on the metasurface and glass substrate. (f) The upconversion luminescence intensities from UCNPs deposited on glass and metasurface, as a function of the excitation power density for the green and red emissions, respectively. (g) Luminescence enhancement factors at emission wavelengths of 540 nm (green) and 660 nm (red) as a function of the 980 nm incident excitation power. (h) Emission lifetime measurements at 540 nm for UCNPs deposited on resonant metasurface supporting Mie resonant mode (green) and on a glass slide (black) under *x*-polarization detection. (i) Time-resolved photoluminescence measurements of UCNPs deposited on a glass slide (black) and a metasurface supporting quasi-BIC mode (red) at an emission wavelength of 660 nm under *y*-polarization detection.

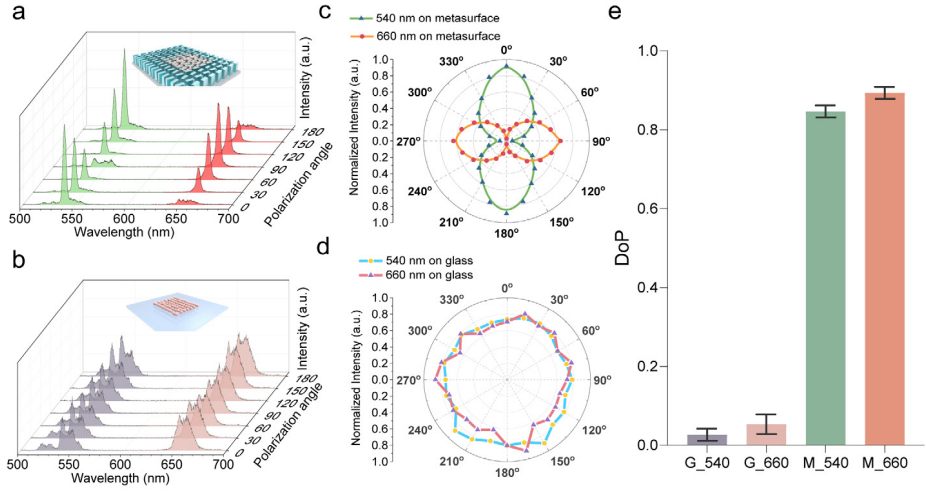

**Figure 4**. (a) Upconversion luminescence spectra of UCNPs coupled to the resonant metasurface, recorded with a rotating analyzer at various angles from 0 to 180° in front of the spectrometer. (b) Upconversion emission spectra from UCNPs on a glass substrate, recorded at different detection polarization angles ranging from 0 to 180°. (c-d) Polar plots illustrating the luminescence intensity of UCNPs coupled to the metasurface (c) and UCNPs on a glass substrate (d) as functions of the emission polarization angle. (e) Degree of polarization (DoP) for upconversion luminescence of UCNPs on both metasurface and glass substrate.